# First-principles study of oxygen vacancy defects in orthorhombic $Hf_{0.5}Zr_{0.5}O_2$/$SiO_2$/Si gate stack

Junshuai Chai,[1,2] Hao Xu,[1,2] Jinjuan Xiang,[1,2] Yuanyuan Zhang,[3] Lixing Zhou,[4] Shujing Zhao,[1,2] Fengbin Tian,[1,2] Jiahui Duan,[1,2] Kai Han,[5] Xiaolei Wang,[1,2, a)] Jun Luo,[1,2] Wenwu Wang,[1,2, a)] and Tianchun Ye[1,2], Yuzheng Guo[6]

**AFFILIATIONS**

[1] Key Laboratory of Microelectronics & Integrated Technology, Institute of Microelectronics, Chinese Academy of Sciences, Beijing 100029, China

[2] College of Microelectronics, University of Chinese Academy of Sciences, Beijing 100049, China

[3] School of Integrated Circuits, Tsinghua University, Beijing 100084, China

[4] Faculty of Information Technology, School of Microelectronics, Beijing University of Technology, Beijing 100124, China

[5] School of Physics and Electronic Information, Weifang University, Weifang 261061, China

[6] School of Electrical Engineering and Automation, Institute of Technological Sciences, Wuhan University, Wuhan 430072, China

[a)] Author to whom correspondence should be addressed: wangxiaolei@ime.ac.cn and wangwenwu@ime.ac.cn

**ABSTRACT**

The gate defect of the ferroelectric HfO$_2$-based Si field-effect transistor (Si FeFET) plays a dominant role in its reliability issue. The first-principles calculations are an effective method for the atomic-scale understanding of gate defects. However, the first-principles study on the defects of FeFET gate stacks, i.e., metal/orthorhombic-Hf$_{0.5}$Zr$_{0.5}$O$_2$/SiO$_x$/Si structure, has not been reported so far. The key challenge is the construction of metal/orthorhombic-Hf$_{0.5}$Zr$_{0.5}$O$_2$/SiO$_x$/Si gate stack models. Here, we use the Hf$_{0.5}$Zr$_{0.5}$O$_2$(130) high-index crystal face as the orthorhombic ferroelectric layer and construct a robust atomic structure of the orthorhombic-Hf$_{0.5}$Zr$_{0.5}$O$_2$/SiO$_2$/Si gate stack without any gap states. Its high structural stability is ascribed to the insulated interface. The calculated band offsets show that this gate structure is of the type-I band alignment. Furthermore, the formation energies and charge transition levels (CTLs) of defects reveal that the oxygen vacancy defects are more favorable to form compared with other defects such as oxygen interstitial and Hf/Zr vacancy, and their CTLs are mainly localized near the Si conduction band minimum and valence band maximum, in agreement with the reported experimental results. The oxygen vacancy defects are responsible for charge trapping/de-trapping behavior in Si FeFET. This work provides an insight into gate defects and paves the way to carry out the first-principles study of ferroelectric HfO$_2$-based Si FeFET.

HfO$_2$-based Si ferroelectric field-effect transistor (FeFET) is perused due to its excellent characteristics, such as complete compatibility with the CMOS process, >10-year retention, low power consumption, fast read/write speed, and scaling ability.[1-8] The limited endurance performance is still a key obstacle to its application.[9-21] The charge trapping/de-trapping behavior with the defects in the gate stacks leads to endurance fatigue.[22-24] However, the physical origin of gate defects in the HfO$_2$-based FeFET is still unknown. Thus the understanding of gate defects is rather essential to improving the endurance.

The first-principles calculations are an effective method for the atomic-scale understanding of gate defects.[25-32] The reported first-principles studies so far mainly include the following three aspects: (i) study of defects in the infinite bulk structure of HfO$_2$-based ferroelectric (FE-HfO$_2$). Wei *et al*. have studied the formation mechanisms of intrinsic point defects under different growth conditions for orthorhombic-Hf$_{0.5}$Zr$_{0.5}$O$_2$ (*o*-HZO) and found that the oxygen vacancy (V$_O$) and oxygen interstitial (O$_i$) are the dominant defects under O-poor and O-rich conditions, respectively.[33] Wei *et al*. have studied the impact of the oxygen defects on ferroelectricity of HZO and found that V$_O$ and Frenkel pair enhances the spontaneous polarization (P$_s$), while O$_i$ largely reduces P$_s$.[34] Zhao *et al*. have studied the effect of group-III dopants (La, Y, Al, and Gd) in HZO ferroelectrics, and found that under stoichiometric doping conditions, La and other group-III dopants increased the formation energy of V$_O$, and La, Al, or Y dopants gathered migrating V$_O$ and passivated the defect states in the bandgap.[35] (ii) study of defects in the TiN/FE-HfO$_2$/TiN capacitor. Zhou *et al*. have studied the impact of interfacial V$_O$ on ferroelectric phase transformation and found that the tetragonal phase would transform into the orthorhombic ferroelectric phase under very few V$_O$ conditions, and as the V$_O$ increases, the monoclinic phase would be dominating.[36] Xue *et al.* have studied the failure mechanism of FE-HfO$_2$ and found that oxygen vacancies tended to be positively charged and further pinned the domain walls, resulting in ferroelectric fatigue.[37] (iii) study of defects in the infinite monoclinic-Hf$_{0.5}$Zr$_{0.5}$O$_2$/orthorhombic-Hf$_{0.5}$Zr$_{0.5}$O$_2$ (*m*-HZO/*o*-HZO) superlattice. Zheng *et al.* have studied the stability of interfacial V$_O$ and found that V$_O$ defects were more easily generated in the *o*-HZO grain with orientation parallel to the electrical field, and as the cycling increased, the formation energies of Vo defects at the *m*-HZO/*o*-HZO interface was lower than that in bulk.[38]

However, the first-principles study on the defects of FeFET gate stacks, i.e., metal/*o*-HZO/SiO$_x$/Si structure, has not been reported so far. The key challenge is the construction of metal/*o*-HZO/SiO$_x$/Si gate stack models. Firstly, the low-index crystal face of *o*-HZO such as (001), (110), (111) and the SiO$_2$/Si (001) generally have a large lattice mismatch, which leads to a large distortion of the orthorhombic ferroelectric layer. Secondly, an insulating interface between *o*-HZO and SiO$_2$ is difficult to achieve because the interfacial dangling bonds are difficultly saturated, which leads to an unstable interface. Although the first-principles investigations of traditional high-$\kappa$ dielectric HfO$_2$/SiO$_2$/Si gate stack have been extensively reported,[30,39-43] the HZO in the FeFET is orthorhombic, which is rather different from the amorphous/monoclinic HfO$_2$ in conventional MOSFET. Thus, it is urgent to construct the metal/*o*-HZO/SiO$_x$/Si structure to understand the gate defects in HfO$_2$-based Si FeFET.

In this work, using the HZO(130) high-index crystal face as the orthorhombic ferroelectric layer, we construct a stable atomic structure of the *o*-HZO/SiO$_2$/Si gate stack. The band offset results reveal the property of its type-I band alignment. Furthermore, by calculating the formation energies and charge transition levels (CTLs), we find that the oxygen vacancies (V$_O$) are more favorable to form compared with other defects such as oxygen interstitial and Hf/Zr vacancy, and the CTLs of V$_O$ are mainly localized near the Si conduction band minimum (CBM) and valence band maximum (VBM). The V$_O$ is responsible for charge trapping/de-trapping behavior in Si FeFET.

The first-principles calculations based on density functional theory were carried out as implemented in the PWmat software with graphics processing unit (GPU) acceleration.[44,45] The electron-ion interaction was described by the projector-augmented-wave (PAW) potential.[46] The Perdew-Burke-Ernzerhof (PBE) of the generalized gradient approximation (GGA) functional was used to optimize the geometric structures.[47] The SG15 norm-conserving pseudopotentials were adopted with an energy cutoff of 50 Ry. The size of the $o$-HZO/SiO$_2$/Si supercell was set to XYZ= 16.38 Å× 5.46 Å× 65 Å with eight HZO atomic layers (~21 Å), four SiO$_2$ atomic layers (~9 Å), and eight Si atomic layers (~9.5 Å). One hydrogen layer was used to passivate the lowest Si dangling bonds in the Si substrate, and the vacuum layer was set at 40 Å in the Z direction to avoid the undesirable image interaction induced by the periodic boundary condition. To obtain an insulating HZO surface, the dangling bonds of half of the topmost O atoms were also passivated by hydrogen atoms. Throughout the calculations, only the one bottom Si layer of the substrate was fixed, while other atoms were fully relaxed. The Brillouin zone (BZ) was sampled with a 2×6×1 Monkhorst-Pack k-point grid. The convergence criteria of electronic self-consistent and ionic relaxation were set to $10^{-4}$ eV and 0.01 eV Å$^{-1}$ for total energy and force, respectively. For the calculation of electronic structures, the Heyd-Scuseria-Ernzerhof (HSE) hybrid functional was adopted to obtain an accurate bandgap.[48] A 1×3×1 Monkhorst k-point mesh and a single gamma point were used in the HSE calculations for the 1×1×1 and the 1×2×1 $o$-HZO/SiO$_2$/Si supercell, respectively. The atomic specific mixing parameters $a_i$ of HSE functional for the different gate stack layers were set separately to obtain their correct band gaps simultaneously, such as $a$=0 for

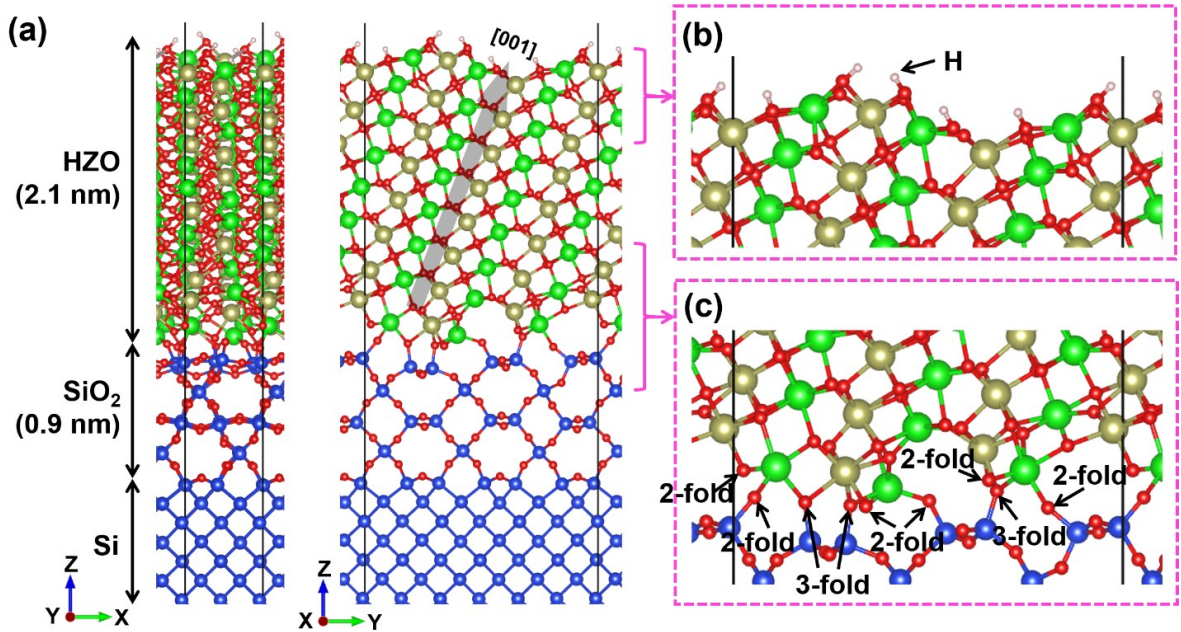

Fig. 1 (a) The side view of *o*-HZO/SiO$_2$/Si gate structure. The black line represents the unit cell. The gray arrow represents the polarization direction along [001]. Brown, Green, red, blue and light pink spheres represent Hf, Zr, O, Si and H atoms, respectively. (b)The magnified images of the *o*-HZO surface. (c) The magnified images of the *o*-HZO/SiO$_2$ interface. 2-fold and 3-fold represent two- and three-coordinated O atoms, respectively.

Hf, Zr, and O atoms of Hf$_{0.5}$Zr$_{0.5}$O$_2$, $a$=0.24 for Si and O atoms of SiO$_2$, and $a$=-0.1 for Si substrate.[49]

We first construct the *o*-HZO/SiO$_2$/Si gate stack models. The fully relaxed atomic structure of the *o*-HZO/SiO$_2$/Si gate stack with 279 atoms is shown in Fig. 1(a). The HZO ferroelectric layer adopts the orthorhombic phase with the space group *Pca*2$_1$. This phase has generally been considered the source of ferroelectric polarization,[50,51] which is attributed to the displacement of three-coordinated O atoms along [001] direction. One can see that the HZO ferroelectric layer no appears the obvious distortion after structural relaxation, indicating the *o*-HZO/SiO$_2$/Si gate structure have ferroelectricity. SiO₂ dielectric layer adopts the *α*-quartz phase with the space group of *P*3₂21. Despite the experimentally deposited HZO ferroelectric layer being polycrystalline with the existence of grain boundaries and the SiO₂ dielectric layer being amorphous, the single crystal orthorhombic HZO and SiO₂ are used here because our stack model is mainly used to research the local structures and local bonding of HZO/SiO₂ interface.[52] To minimize the strain introduced by the interface, a supercell with the interface match relation of *o*-HZO(130)‖SiO₂(110)‖Si(001) is built along the Z-oriented Si substrate. Along the X-axis direction of this supercell, the lattice length $X_{Si}$ (5.46 Å) corresponds to the $X_{SiO_2}$ (5.03 Å) and $X_{HZO}$ (5.31 Å) with the lattice mismatch of 7.88% and 2.75%, respectively, and along the Y-axis direction, the lattice length $Y_{Si}$ (5.46×3=16.38 Å) corresponds to the $Y_{SiO_2}$ (5.51×3=16.53 Å) and $Y_{HZO}$ (16.10 Å) with a small lattice mismatch of -0.09% and 1.71%, respectively. Here, we have also considered the *o*-HZO with low-index crystal faces and the SiO₂/Si matching, but the corresponding HZO ferroelectric layers have appeared the severe structural distortion after structural relaxation due to the large lattice mismatches and unstable interfaces (not shown here).

Fig. 1(c) shows the magnified images of the *o*-HZO/SiO₂ interface of our constructed *o*-HZO/SiO₂/Si gate structure. One can see that the dominated bonds are Hf(Zr)-O-Si bonding, which has been confirmed more favorable than other possible bonding types such as Hf(Zr)-Si bonding. The bond length of Hf(Zr)-O and Si-O are 1.96~2.32 Å and 1.67 Å, respectively, which are consistent with the bond length in bulk *o*-HZO (2.04~2.28 Å) and SiO₂ (1.63Å). There are nine interfacial O atoms in the unit cell: six two-coordinated and three three-coordinated O atoms. The two-coordinated O atoms contain two bonding types: (i) O atom bonds to one Hf and Zr atom; (ii) O atom bonds to one Zr atom from HZO ferroelectric side and one Si atom from SiO₂ dielectric side. Each three-coordinated O atom bonds one Hf and Zr, and one Si atom. Here, although these interfacial Hf and Zr atoms are under-coordinated compared with the bulk phase, they achieve closed-shell because all their valence electrons have been transferred to O atoms around them to fully saturate the dangling bonds of O atoms. This ensures that the $o$-HZO/SiO$_2$ interface satisfies the electronic counting rule (ECR) and shows an insulating property. For the SiO$_2$/Si interface, the oxygen-terminated SiO$_2$ surface is used to contact the Si(001) substrate, similar to the previous report.[53] This interface also obeys the ECR and presents an insulating property. The above insulating interfaces are the key to guaranteeing high structural stability and accurately calculating the band alignment of the interface for the $o$-HZO/SiO$_2$/Si gate structure.

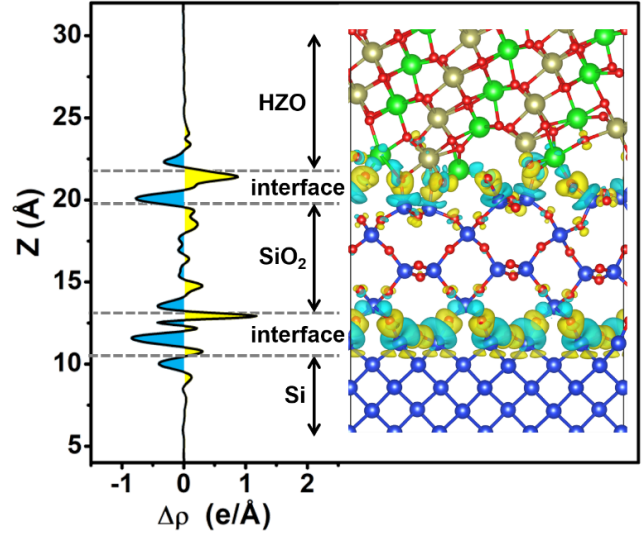

Fig. 2 The three-dimensional (right panel) and the corresponding planar-averaged (left panel) charge density difference along the Z-direction for the $o$-HZO/SiO$_2$/Si gate structure. Yellow and blue parts represent the charge accumulation and depletion, respectively. The isovalue is 0.001 e/Å$^3$.

To evaluate the interfacial stability of the $o$-HZO/SiO$_2$/Si gate structure, the charge density difference (CDD) is calculated. The CDD is defined as $\Delta\rho(r)=\rho[\text{HZO/SiO}_2\text{/Si}]-\rho[\text{HZO}]-\rho[\text{SiO}_2]-\rho[\text{Si sub}]$, where $\rho[\text{HZO/SiO}_2\text{/Si}]$ is the charge density of $o$-HZO/SiO$_2$/Si gate structure, and $ρ$[HZO], $ρ$[SiO$_2$] and $ρ$[Si sub] are the charge density of isolated *o*-HZO ferroelectric layer, SiO$_2$ dielectric layer, and Si substrate, respectively. The calculated three-dimensional CDD and the corresponding planar-averaged CDD along the Z-direction are shown in Fig. 2. For the *o*-HZO/SiO$_2$ interface, a clear charge depletion region (blue part) appears around interfacial Hf(Zr) and Si atoms, while a charge accumulation region (yellow part) appears around interfacial O atoms. This means that electrons transfer from interfacial Hf(Zr) and Si atoms to the interfacial O atoms. Thus, the electric dipoles form at the *o*-HZO/SiO$_2$ interface, which reflects the interfacial polarization effect. For the SiO$_2$/Si interface, a region with larger charge depletion and accumulation appears, respectively, around interfacial Si and O atoms compared with *o*-HZO/SiO$_2$ interface. This means that there exists a stronger charge transfer, reflecting a stronger interfacial polarization effect.

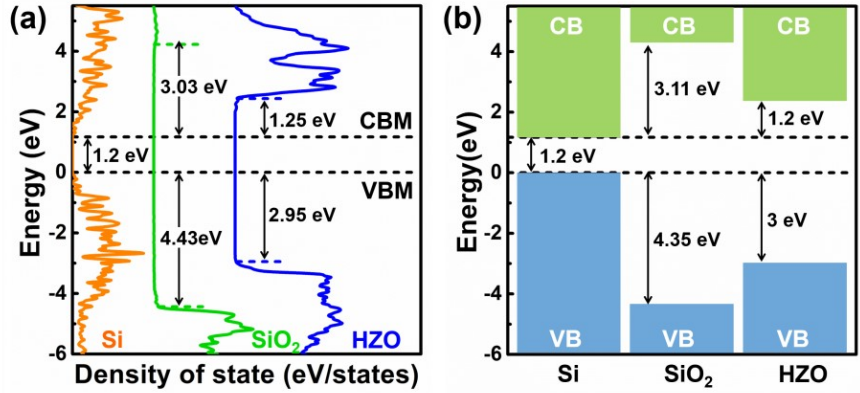

Fig. 3 The band alignment of interface based on the LDOS (a) and electrostatic potential alignment method (b) for the *o*-HZO/SiO$_2$/Si gate structure. The calculated bandgaps of *o*-HZO, SiO$_2$ and Si are 5.4, 8.66 and 1.2 eV, respectively. CB and VB represent the conduction band and valence band, respectively.

We next investigate the energy band alignment of the *o*-HZO/SiO$_2$/Si gate structure. Fig. 3(a) shows the calculated local density of states (LDOS) of bulk-like Si, SiO$_2$, and *o*-HZO far from the interface region. The *o*-HZO/SiO$_2$/Si gate structure does not have any gap states, which means that the system satisfies the ECR. The bandgap of the *o*-HZO ferroelectric layer, SiO$_2$ dielectric layer, and Si substrate are 5.4, 8.66, and 1.2 eV, respectively, which well reproduces the experimental bandgap values.[54] The valence band maximum (VBM) and conduction band minimum (CBM) of the Si substrate are higher and lower than that of the oxide layer (*o*-HZO and SiO$_2$), respectively, indicating that *o*-HZO/Si and SiO$_2$/Si interfaces are of type-I band alignment. A similar phenomenon appears at *o*-HZO/SiO$_2$ interface, also reflecting the type-I band alignment.

Since the band offset evaluated based on LDOS calculation is usually rough, the electrostatic potential alignment method[55] is further used to calculate the band alignment accurately. In this method, the valance band offset (VBO) can be evaluated by the expression $E_{VBO}=E_{VBM}$(bulk Si)-$E_{VBM}$(bulk oxides)-$\Delta V$, where the $E_{VBM}$(bulk Si) and $E_{VBM}$(bulk oxides) are the valance band maximum of bulk Si and oxides (*o*-HZO or SiO$_2$), respectively, and the $\Delta V$ is the difference of electrostatic potential between the Si substrate and oxides layer. The conduction band offset (CBO) can be evaluated by $E_{CBO}=E_{VBO}-[E_g$(bulk Si)-$E_g$(bulk oxides)], where $E_g$(bulk Si) and $E_g$(bulk oxides) are the bandgaps of bulk Si and oxides, respectively. The calculated band alignment (see Fig. 3(b)) shows that the VBO of SiO$_2$/Si and *o*-HZO/Si interfaces are 4.35 and 3 eV, respectively, and the corresponding CBO are 3.11 and 1.2 eV.

We finally investigate the charge transition levels (CTLs) of defects (oxygen vacancy and interstitial, and metal atom vacancy) to evaluate the charge trapping/de-trapping behavior in the *o*-HZO/SiO$_2$/Si gate structure. The key quantity that determines the CTLs is the defect formation energy, which can reflect the stability of defects. The defect formation energy is defined as $\Delta H(D_i,$

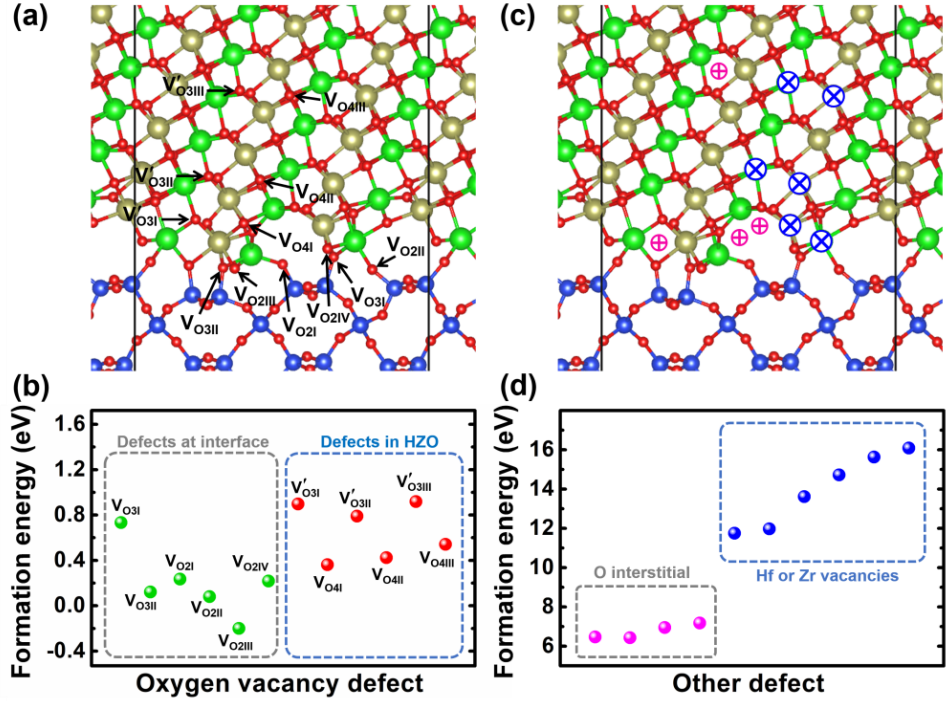

Fig. 4 (a) The oxygen vacancy (V$_O$) defects with different positions for the *o*-HZO/SiO$_2$/Si gate structure. V$_{O3I}$, V$_{O3II}$, V$_{O2I}$, V$_{O2II}$, V$_{O2III}$, and V$_{O2IV}$ represent the three- and two-coordinated V$_O$ defects at *o*-HZO/SiO$_2$ interface, respectively. V′$_{O3I}$, V′$_{O3II}$, V′$_{O3III}$, V$_{O4I}$, V$_{O4II}$ and V$_{O4III}$ represent the three- and four-coordinated V$_O$ defects in *o*-HZO, respectively. (b) The formation energy of V$_O$ defects with the neutral state. (c) The oxygen interstitial (markded as pink '⊕') and Hf/Zr vacancy (marked as blue '⊗') defects with different positions. (d) The formation energy of O interstitial and Hf/Zr vacancy defects with the neutral state.

$q$)=$E(D_i, q)$-$E(host)$+$(E_i+\mu_i)$+$q[E_V(host)+E_F+\Delta V]$,[56] where $E(D_i, q)$ and $E(host)$ are the total energy of the gate structure with and without $D_i$ defects and net charge state $q$, respectively. $\mu_i$ is the chemical potential of element *i* referenced to the energy of its elemental phase $E_i$ (per atom). $E_V$ is the VBM of gate stack with perfect structure and is used as the reference level of electron chemical potential. $E_F$ is the Fermi level relative to VBM. $\Delta V$ is the difference of electric potential between structures with and without defect at a position (denoted as $R$) far away from the defect, expressed as $\Delta V=V(D_i, q, R)-V(host, R)$. Then the CTLs can be defined as the Fermi-level $E_F$ position at which the formation energy $\Delta H(D_i, q)$ of defect $D_i$ with the charge state $q$ is equal to that of the same defect with another charge $q'$, expressed as $\varepsilon(q/q')=[\Delta H(D_i, q)-\Delta H(D_i, q')]/(q-q')$.

Fig. 4 (a) and (b) display the calculated formation energies of oxygen vacancy ($V_O$) defects with the neutral state at different physical positions, including at the $o$-HZO/SiO$_2$ interface (three-coordinated $V_{O3I}$ and $V_{O3II}$, and two-coordinated $V_{O2I}$, $V_{O2II}$, $V_{O2III}$, and $V_{O2IV}$) and in $o$-HZO (three-coordinated $V'_{O3I}$, $V'_{O3II}$, and $V'_{O3III}$, and four-coordinated $V_{O4I}$, $V_{O4II}$, and $V_{O4III}$). The formation energies of $V_O$ at the $o$-HZO/SiO$_2$ interface, especially $V_{O2III}$, are generally lower than that of $V_O$ in $o$-HZO, indicating that the interfacial $V_O$ defects are more favorable than other $V_O$ for $o$-HZO/SiO$_2$/Si gate structure because the Hf(Zr)-O-Si bonds at the $o$-HZO/SiO$_2$ interface are weaker than the Hf(Zr)-O bonds in $o$-HZO. Note that for defects at $o$-HZO/SiO$_2$ interface, although both $V_{O3I}$ and $V_{O3II}$ defects are the three-coordinated, the formation energy of the former is larger than the latter. This is ascribed to the different coordination of Hf and Zr atoms around them. The Hf and Zr atoms around $V_{O3II}$ are seven- and five-coordinated, respectively, while both of them around $V_{O3I}$ are six-coordinated. A similar phenomenon also appears between two-coordinated $V_{O2I}$ and $V_{O2II}$. So is between $V_{O2III}$ and $V_{O2IV}$ defects. For defects in $o$-HZO, the formation energies of three-coordinated $V_O$ defects are ~0.4 eV larger than that of four-coordinated $V_O$, which are consistent with the previous first-principles studies on HfO$_2$-based bulk structure.[33,34,57-59] In addition, Fig. 4 (c) and (d) display the calculated formation energies of oxygen interstitial and Hf/Zr vacancy. It shows that their formation energies are much larger than that of $V_O$ and consequently they are rather difficult to form. Thus we focus on $V_O$ defects in the following section.

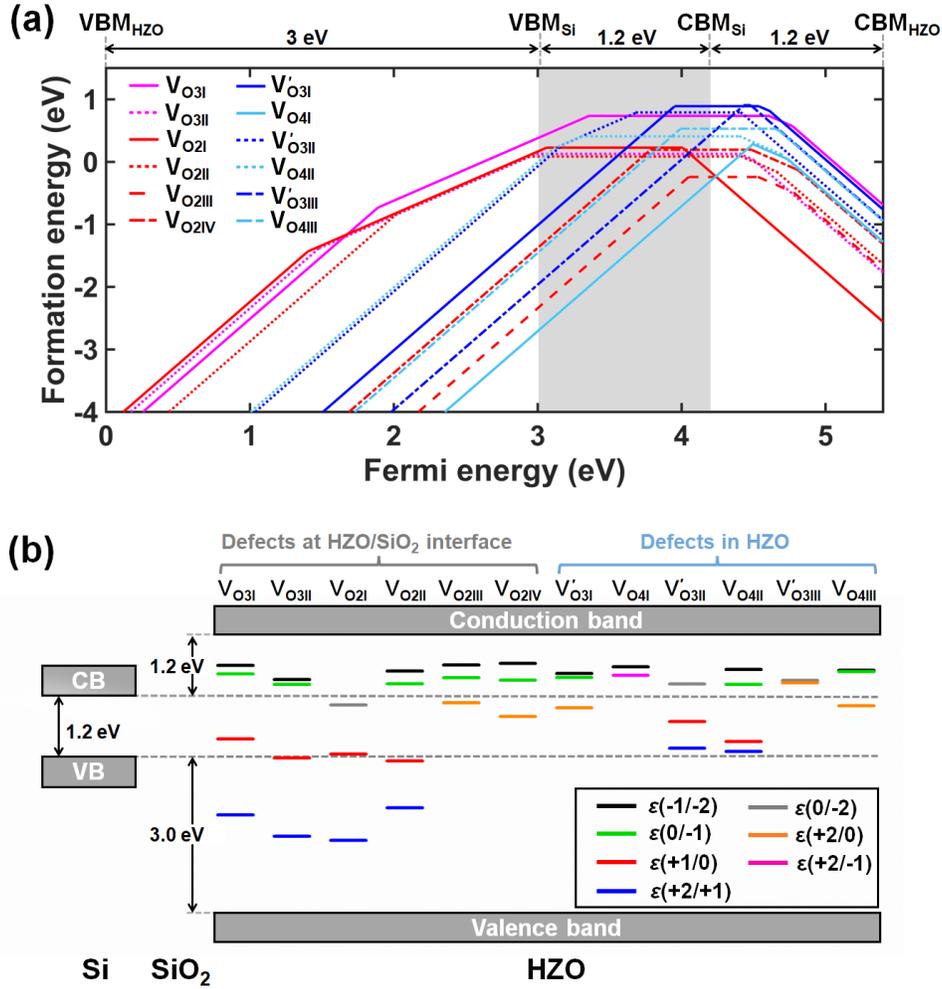

Fig. 5 (a) The formation energies of $V_O$ defects as a function of Fermi level $E_F$ for the $o$-HZO/SiO$_2$/Si gate structure. The grey region represents the Si bandgap, and the SiO$_2$ region is not shown. (b) The charge transition levels (CTLs) of $V_O$ defects at $o$-HZO/SiO$_2$ interface and in $o$-HZO. Black, green, red, blue, grey, orange and pink lines represent the $\varepsilon(-1/-2)$, $\varepsilon(0/-1)$, $\varepsilon(+1/0)$, $\varepsilon(+2/+1)$, $\varepsilon(0/-2)$, $\varepsilon(+2/0)$, and $\varepsilon(+2/-1)$ transition levels, respectively.

The formation energies of various $V_O$ defects as a function of Fermi level $E_F$ for $o$-HZO/SiO$_2$/Si gate structure are shown in Fig. 5(a). The grey region represents the Si bandgap, and the SiO$_2$ region is not shown. The CTLs are extracted and plotted in Fig. 5(b) and the corresponding values are listed in Table I. From Fig. 5(b), one can see that all kinds of $V_O$ defects at the $o$-HZO/SiO$_2$ interface and in $o$-HZO have CTLs near the Si conductance band, while the following kinds of $V_{O3I}$, $V_{O3II}$, $V_{O2I}$, $V_{O2II}$, $V'_{O3II}$, and $V_{O4II}$ have CTLs near the Si valence band. The experimentally extracted defect levels were reported to be ~0.5 eV above Si CBM.[60-62] Thus our calculation results are well consistent with the experimental data. This indicates that the defect responsible for charge trapping/de-trapping in Si FeFET is oxygen vacancy. Thus, the passivation of $V_O$ is beneficial to the improvement of endurance properties for HfO$_2$-based Si FeFET.

Table I The value of CTLs of $V_O$ defects at $o$-HZO/SiO$_2$ interface and in $o$-HZO for the $o$-HZO/SiO$_2$/Si gate structure.

| $\varepsilon(q/q')$ | Oxygen vacancy defect | | | | | | | | | | | |
|---|---|---|---|---|---|---|---|---|---|---|---|---|
| | Defect at $o$-HZO/SiO$_2$ interface | | | | | | Defect in $o$-HZO | | | | | |
| | $V_{O3I}$ | $V_{O3II}$ | $V_{O2I}$ | $V_{O2II}$ | $V_{O2III}$ | $V_{O2IV}$ | $V'_{O3I}$ | $V_{O4I}$ | $V'_{O3II}$ | $V_{O4II}$ | $V'_{O3III}$ | $V_{O4III}$ |
| $\varepsilon(-1/-2)$ | 4.77 | 4.50 | | 4.66 | 4.78 | 4.81 | 4.62 | 4.74 | | 4.69 | | 4.68 |
| $\varepsilon(0/-1)$ | 4.61 | 4.41 | | 4.42 | 4.53 | 4.48 | 4.54 | | | 4.40 | | 4.65 |
| $\varepsilon(+1/0)$ | 3.36 | 2.99 | 3.06 | 2.93 | | | | | 3.69 | 3.30 | | |
| $\varepsilon(+2/+1)$ | 1.89 | 1.48 | 1.40 | 2.03 | | | | | 3.17 | 3.11 | | |
| $\varepsilon(0/-2)$ | | | 4.00 | | | | | | 4.41 | | 4.48 | |
| $\varepsilon(+2/0)$ | | | | | 4.05 | 3.78 | 3.95 | | | | 4.43 | 3.99 |
| $\varepsilon(+2/-1)$ | | | | | | | | 4.54 | | | | |

  In conclusion, we construct a stable atomic structure of the ferroelectric $o$-HZO/SiO$_2$/Si gate stack by using the $o$-HZO(130) crystal face as the ferroelectric layer. The insulating interfaces guarantee its high structural stability. This gate structure is of type-I band alignment. Furthermore, the calculated formation energies and CTLs reveal that the oxygen vacancy defects are more favorable to form compared with other defects such as oxygen interstitial and Hf/Zr vacancy, and their CTLs mainly are localized near the Si CBM and VBM. The oxygen vacancy defects are responsible for charge trapping/de-trapping in Si FeFET. This work provides an insight into gate defects and paves the way to carry out the first-principles study of ferroelectric $HfO_2$-based Si FeFET.

This work was supported in part by the National Natural Science Foundation of China under Grant 61904199 and Grant 61904193, in part by the China Postdoctoral Science Foundation under Grant E1BSH1X001, and in part by the Natural Science Foundation of Beijing Municipality under Grant 4214079.

## DATA AVAILABILITY

The data that supports the findings of this study are available from the corresponding author upon reasonable request.